\documentclass[12pt]{article}
\usepackage{euscript,amsmath,amssymb,latexsym}

\newcommand{\be}{\begin{equation}}
\newcommand{\ee}{\end{equation}}
\newcommand{\bea}{\begin{eqnarray}}
\newcommand{\eea}{\end{eqnarray}}

\begin{document}
\title{\bf
\vspace{1cm}Quantization of the Riemann Zeta-Function and Cosmology}
\author{
I. Ya. Aref'eva and I.V. Volovich
 \\
 ~~
 \\
{\it  Steklov Mathematical Institute}
\\ {\it Gubkin St.8, 119991 Moscow, Russia}}

\date {~}
\maketitle
\begin{abstract} Quantization of the Riemann zeta-function is proposed.
We treat the Riemann
zeta-function as a symbol of a pseudodifferential operator and study the corresponding
classical and quantum field theories. This approach is motivated by the theory of
$p$-adic strings and by recent works on stringy cosmological models. We show that the
Lagrangian for the zeta-function field is equivalent to the sum of the Klein-Gordon
Lagrangians with masses defined by the zeros of the Riemann zeta-function. Quantization
of the mathematics of
 Fermat-Wiles and the Langlands program   is indicated. The Beilinson conjectures
on the values of $L$-functions of motives are interpreted as dealing
with the cosmological constant problem. Possible cosmological
applications of the zeta-function field theory are  discussed.

\end{abstract}

\newpage
\section{Introduction}

Recent astrophysical data require  rather exotic field models that can violate the null
energy condition (see \cite{AV} and refs. therein). The linearized  equation for the
field $\phi$ has the form
$$
F(\Box)\phi =0\,,
$$
where $\Box$ is the d'Alembert operator and $F$ is an analytic
function.
 Stringy models provide a possible candidate for this
 type of models. In
particular, in this context $p$-adic string models
\cite{p-adic,Man,Var} have been considered. $p$-Adic cosmological
stringy models are supposed to incorporate essential features of
usual string models \cite{IA,Calcagni,Cline}.

An advantage of the $p$-adic string is that it can be described by an effective field
theory including just one scalar field. In this effective field theory the kinetic term
contains an operator \be \label{p-adic-kin} p^{\Box}\,, \ee where $p$ is a prime
number. In
the spirit of the adelic approach \cite{Man, adelic} one could try an adelic product \be
\label{adel-kin} \prod_pp^{\Box}\,, \ee
 but (\ref{adel-kin}) has no obvious
mathematical meaning. There is the celebrated adelic  Euler
formula for the zeta-function \cite{anal-number-theory}
$$
\zeta(s)=\prod_p(1-p^{-s})^{-1}\,,
$$
therefore a natural operator should be related with the Riemann zeta-function $\zeta
(s)$. Due to the Riemann hypothesis on zeros of the zeta-function the most interesting
operator is the following  pseudodifferential operator, \be \label{I2}
\zeta(\frac{1}{2}+i\Box)\,. \ee
We call this operator the {\it quantum zeta-function}.
We consider also quantization
of the Riemann $\xi$-function and various $L$-functions.
Euclidean version of the quantum zeta-function
is $\zeta(\frac{1}{2}+i\Delta)$ where $\Delta$ is
the Laplace operator.

One can wonder what is special
about the zeta-function and why we do not consider an arbitrary analytic function?
One of
the reasons  is that there is a universality of the Riemann zeta-function. Any
analytic function can be approximated by the Riemann zeta-function. More precisely,
  any nonvanishing analytic
function can be approximated uniformly by certain purely imaginary shifts of the
zeta-function in the critical strip (Voronin`s theorem \cite{anal-number-theory}). Remind
that the Riemann zeta-function has  appeared in different problems in mathematical and
theoretical physics, see \cite{adelic,zeta-mp}.

Under the assumption that the Riemann hypothesis is true we show that the
Lagrangian for the zeta-function field is equivalent to the sum of the Klein-Gordon
Lagrangians with masses defined by the zeros of the Riemann zeta-function
at the critical line. If the Riemann hypothesis is not true then the mass spectrum
of the field theory is different.

An approach to the derivation of the mass spectrum of
elementary particles by using solutions of the Klein-Gordon
equation with finite action is considered in \cite{KozVol}.

We can quantize  not only the Riemann zeta-function but also
more general $L$-functions. Quantization
of the mathematics of
 Fermat-Wiles \cite{Wil} and the Langlands program \cite{Lan}  is indicated.

The paper is organized as follows. In Section 2 we collect
an information about the Riemann zeta-function. In
Section 3 we present a classical field theory
with the zeta-function kinetic term and discuss
 solutions   of the model in Minkowski space.
  Then we study dynamics in the Friedmann metric in some approximation and discuss
cosmological properties of the constructed solutions. In Section 4 there are few comments
about the corresponding quantum theory and in Sections 5 and 6 we consider  modifications
of the theory where instead of the zeta-function kinetic term  the Riemann-Siegel
function or $L$-function are taken.

\section{Riemann Zeta-Function}

Here we collect some information about the Riemann zeta-function
which we shall use in the next section to study the zeta-function
field theory.
The Riemann zeta-function is defined as \be \label{R1}
\zeta(s)=\sum_{n=1}^{\infty}\frac{1}{n^s},~~s=\sigma+i\tau,~\sigma
>1 \,,\ee
and there is an Euler adelic representation

\be \label{R2} \zeta(s)=\prod_p(1-p^{-s})^{-1} \ee (we write $\tau$
instead of $t$ which is usually used for the imaginary part of $s$
since we shall use $t$ as a  time variable). The zeta-function
admits an analytic continuation to the whole complex plane $s$
except the point $s=1$ where it has a simple pole.
It goes as follows. We define the theta-function
\be \label{R2t} \theta (x)=\sum_{n=-\infty}^{\infty}
e^{-\pi n^2 x},~~x>0 \,,\ee
which satisfies the functional equation
\be
\label{R2t1} \theta (x)=
\frac{1}{\sqrt{x}}\theta(\frac{1}{x})
\ee
and is an example of a modular form. Roughly speaking the
Riemann zeta-function $\zeta(s)$ is the Mellin transform of
the theta-function $\theta (x).$

The "Langlands philosophy" \cite{Lan} says that all reasonable
generalizations of the Riemann zeta-function are related with
modular forms, see Sect.6.

One introduces the Riemann $\xi$-function \be \label{R3}
\xi(s)=\frac{s(s-1)}{2}\pi^{-\frac{s}{2}}\Gamma(\frac{s}{2})\zeta(s) \ee
and obtains the
expression \be \label{R3t} \xi (s)=\frac{s(s-1)}{2}\int_{1}^{\infty}(x^{\frac{s}{2}-1}+
x^{-\frac{s}{2}-\frac{1}{2}})\omega (x)dx+\frac{1}{2}\,,
\ee
where
\be \label{R21t} \omega
(x)=\frac{1}{2}(\theta (x)-1)=\sum_{n=1}^{\infty} e^{-\pi n^2 x}\,.
\ee
The right-hand side
in (\ref{R3}) is meaningful for all values $s\in \mathbb{C},$ since the integral
converges by virtue of the exponential decay of $\omega (x).$ The gamma-function
$\Gamma(s/2)$ never vanishes. The Riemann $\xi$-function $ \xi(s)$ is an entire function.

The zeros of $\zeta (s)$ lie in the critical strip $0<\sigma< 1$
with the exception of the "trivial zeros" at  $s=-2,-4,-6,...$. They
are situated symmetrically about the real axis $\tau =0$ and the
critical line $\sigma=1/2.$ If $\rho$ is a nontrivial zero then
$\bar{\rho},1-\rho$ and $1-\bar{\rho}$ are also zeros. If $N(T)$ is
the number of zeros in the critical strip,
\be
\label{R6n}
N(T)=\#\{\rho=\beta+i\gamma: 0\leq\beta\leq 1,~0\leq\gamma\leq T\}\,,
\ee
then
\be
\label{R6n1}
N(T)=\frac{T}{2\pi}\log \frac{T}{2\pi}-\frac{T}{2\pi}+O(\log T)\,.
\ee
The Riemann hypothesis asserts that all nontrivial zeros $\rho$ lie
at the critical line: $\rho=\frac{1}{2}+i\tau$. There is a
conjecture that all zeros are simple. The first few zeros occur
approximately at the following values: $\tau = 14.1,~
21.0,~25.0,~30.4,~32.9$. The corresponding negative values are also
zeros.

The zeros of the $\xi$-function are the same as the nontrivial zeros
of the $\zeta$-function. It is known that
 $\xi(\frac{1}{2}+i\tau)$ is real for real $\tau$ and
 is bounded.
 Locating zeros on the critical line of the (complex) zeta
 function reduces to locating
 zeros on the real line of the real function $\xi(\frac{1}{2}+i\tau)$.

 There is the functional equation \be \label{R5} \xi(s)=\xi(1-s) \ee
and the Hadamard representation for the $\xi$-function \be
\label{R6}
\xi(s)=\frac{1}{2}e^{as}\prod_{\rho}(1-\frac{s}{\rho})e^{s/\rho}\,.
\ee
Here $\rho$ are nontrivial zeros of the zeta-function and
\be
\label{R5a} a=-\frac{1}{2}\gamma-1+\frac{1}{2}\log 4\pi
\ee
where
$\gamma$ is Euler's constant. The series
\be
\label{R5c}
\sum_{\rho}
\frac{1}{|\rho|^{1+\epsilon}}
\ee
converges for any $\epsilon >0$ but
diverges if $\epsilon =0.$

\section{Zeta-Function Classical Field Theory}
\subsection{Minkowski space}
If $F(\tau)$ is  a function of a real variable $\tau$ then we define
a pseudodifferential operator $F({\Box})$ \cite{Hor} by using the Fourier
transform
\be \label{Z2}
 F({\Box})\phi (x)=\int e^{ixk}F(k^2)\tilde{\phi}(k)dk  \,.
 \ee
Here $\Box$ is the  d'Alembert operator \be \label{Z1a}
{\Box}=-\frac{\partial^2}{\partial x_0^2}+\frac{\partial^2}{\partial
x_1^2}+...+\frac{\partial^2}{\partial x_{d-1}^2}\,, \ee $\phi(x)$ is a
function from $x\in \mathbb{R}^d$, $\tilde{\phi}(k)$ is the Fourier
transform and $k^2=k_0^2-k_1^2-...-k_{d-1}^2$. We assume that the
integral (\ref{Z2}) converges, see \cite{VlaVolY} for a
consideration of one-dimensional $p$-adic field equations.

One can introduce a natural field theory related with the real
valued function $F(\tau)=\xi (\frac{1}{2}+i\tau)$  defined by means
of the zeta-function. We consider the following Lagrangian
\be
\label{Z1L} {\cal L} = \phi \xi (\frac{1}{2}+i{\Box})\phi \,.
\ee
The integral
\be \label{Z2xi}
 \xi (\frac{1}{2}+i{\Box})\phi (x)=\int e^{ixk}
 \xi (\frac{1}{2}+ik^2)\tilde{\phi}(k)dk
 \ee
 converges if $\phi(x)$ is a
 decreasing function
 since $\xi(\frac{1}{2}+i\tau)$ is bounded.

By using (\ref{R3t}) the quantized $\xi$-function  can be expressed
as \be \label{R5c1} \xi(\frac{1}{2}+i\Box)=\frac{1}{2}-(\Box^2
+\frac{1}{4}) \int_{1}^{\infty}x^{-\frac{3}{4}}\cos
[\frac{\Box}{2}\log x ]\omega (x)dx \,.\ee

 The operator
 $\xi (\frac{1}{2}+i{\Box})$ (or $\zeta (\frac{1}{2}+i{\Box})$)
 is the first quantization the Riemann zeta-function. Similarly one
 can define operators $\zeta (\sigma+i{\Box})$
 and $\zeta (\sigma+i{\Delta})$ where $\Delta$ is the Laplace
 operator. We will obtain the second quantization of the Riemann
 zeta-function when we quantize the field $\phi (x).$

Let us prove the following
\\
{\bf Proposition.} {\it The Lagrangian (\ref{Z1L}) is equivalent to the
following Lagrangian}
\be
\label{Z1aa}
 {\cal L}^\prime =
\sum_{\epsilon, n} \eta_{\epsilon n}\psi_{\epsilon n}(\Box +\epsilon
m_n^2)\psi_{\epsilon n}\,,
\ee
{\it where the notations are defined below.}

Let \be \label{Z3} \rho_n= \frac{1}{2}+im_n^2,~~\bar{\rho}_n=
\frac{1}{2}-im_n^2,~~m_n>0,~~n=1,2,... \ee be the zeros at the
critical line.

We shall show that the zeros $m_n^2$ of the Riemann zeta-function
become the masses of elementary particles in the Klein-Gordon
equation.

From the Hadamard representation (\ref{R6}) we get \be \label{Z1h}
\xi (\frac{1}{2}+i\tau)= \frac{C}{2}\prod_{n=1}^{\infty}
(1-\frac{\tau^2}{m_n^4}) \ee because \be \label{Z1hh}
-\frac{1}{2}\gamma-1+\frac{1}{2}\log 4\pi
+\sum_{n=1}^{\infty}\frac{1}{(1/4)+m_n^4}=0 \,.\ee Here \be \label{Z2h}
C= \prod_n \frac{1}{1+\frac{1}{4m_n^4}} \,.\ee Remind that \be
\label{Z2hs} m_n^2\sim2\pi n/\log n,~~~n\to\infty\,.\ee It will be
convenient to write the formula (\ref{Z1h}) in the form \be
\label{Z1c} \xi (\frac{1}{2}+i\tau)= \frac{C}{2}\prod_{\epsilon, n}
(1+\frac{\tau}{\epsilon m_n^2})\,, \ee where $\epsilon =\pm 1$ and a
regularization is assumed. Then our Lagrangian (\ref{Z1L}) takes the
form \be \label{Z1w} {\cal L} = \phi \xi (\frac{1}{2}+i{\Box})\phi =
\frac{C}{2}\phi \prod_{\epsilon, n} (1+\frac{\Box}{\epsilon
m_n^2})\phi \,.\ee
Now if we define the fields $\psi_{\epsilon n}$ as \be
\label{psi-phi} \psi_{\epsilon_0 n_0}~=
~\frac{C}{2m_n^2}~\prod_{\epsilon n\neq \epsilon_0 n_0}~\left(1
+\frac{\Box}{\epsilon m_n^2}\right) \phi \,,\ee and the real constants
$\eta_{\epsilon n}$ by \be \label{eta} \eta_{\epsilon n}~=
~\frac{1}{i\xi ^\prime (\frac12-i\epsilon m_n^2)}\,, \ee then it is
straightforward to see  that the Lagrangians (\ref{Z1L}) and
(\ref{Z1aa}) are equivalent. The proposition is proved.

Similar but different Lagrangians are considered in \cite{AV,PaisU}.

\subsection{The state parameter}

The energy and pressure corresponding to the individual space homogeneous
 (depending only on time $x_0=t$) fields
$\psi=\psi_{\epsilon n}(t)$  in (\ref{Z1aa}) are \cite{BogSchir}
\be \label{E-psi} E=\eta
(~\dot{\psi}^2-\epsilon m^2\psi^2),~~~ P=\eta (~\dot{\psi}^2+\epsilon m^2\psi^2)\,.
\ee
Solutions of the equation of motion for $\psi$
\be
\label{eom}
\ddot{\psi}-\epsilon ~m^2\psi=0
\ee
are
\bea
\label{p}
\psi&=&Ae^{mt}+Be^{-mt},~~~~~~~\epsilon =+1\,,\\
\label{m} \psi&=&F\sin(m(t-t_0)),~~~~~\epsilon =-1\,, \eea where
$A,B,F$ and $t_0$ are real constants. The
state parameter $w=P/E$ corresponding to the solution with
$\epsilon=+1$ is
$$
w=-\frac{1}{2AB}(A^2e^{2mt}+B^2e^{-2mt})\,,
$$
and with $\epsilon=-1$  is
$$
w=\cos(2m(t-t_0)) \,.
$$
We see that if $\epsilon =+1 $ the state parameter $w$ can be
positive or negative depending on the value of $AB$. In the case
$\epsilon =-1$ the state parameter oscillates. The behavior of the
state parameter $w$ is important for cosmological applications, see
for example \cite{AJKV}.

\subsection{Zeta-Function Field Theory in Curved Space-Time}
We couple the zeta-function  field with gravity. We consider the
Lagrangian \be \label{ZC1} {\cal L}=R+\phi F(\Box)\phi +\Lambda  \,.\ee
Here $R$ is the scalar curvature of the metric tensor $g_{\mu\nu}$
with the signature $(-+++)$, $\Lambda$
 is a (cosmological)
 constant and $\Box$ is the d'Alembertian
 $$
 \Box =\frac{1}{\sqrt{-g}}\partial_{\mu}(\sqrt{-g}g^{\mu\nu})\partial_{\nu} \,.
$$
Such a Lagrangian for various choices of the function $F(\Box)$ was
considered in \cite{AV,AJKV}. For the zeta-function field theory,
when $F(\Box)=\xi (\frac{1}{2}+i\Box)$ there is a problem of how to
define rigorously the operator $\xi (\frac{1}{2}+i\Box)$ in the
curved space-time. One can prove, by using a regularized Hadamard
product, similarly to the previous proposition, that the Lagrangian
(\ref{ZC1}) is equivalent to the following Lagrangian \be
\label{ZC2}
 {\cal L}^\prime =R+
\sum_{\epsilon, n} \eta_{\epsilon n}\psi_{\epsilon n}(\Box +\epsilon
m_n^2)\psi_{\epsilon n}+\Lambda \,.\ee Then one can study various
cosmological solutions along the lines of \cite{AJKV}. As compare to
models considered in \cite{AJKV} in (\ref{ZC2}) there is no an extra
exponent of an entire function. This permits to find  deformations
of the model (\ref{ZC1}) that admits exact solutions.

For a homogeneous, isotropic and the spatially flat Universe with
the metric
\begin{equation}
\label{mFr} ds^2={}-dt^2+a^2(t)\left(dx_1^2+dx_2^2+dx_3^2\right)\,,
\end{equation}
we obtain the Friedmann equations (we set $\Lambda =0$)
\begin{equation}
\label{mFr2} H^2=\frac{1}{6}\sum_{\epsilon,n}\eta_{\epsilon n}
(\dot{\psi}_{\epsilon n}^2-\epsilon m_n^2\psi_{\epsilon n}^2)\,,
\end{equation}
$$
\dot{H}=-\frac{1}{2}\sum_{\epsilon,n}\eta_{\epsilon n} \dot{\psi}_{\epsilon n}^2\,,
$$
$$
\ddot{\psi}_{\epsilon n}+3H\dot{\psi}_{\epsilon n}-\epsilon
m_n^2\psi_{\epsilon n}=0\,,
$$
where the Hubble parameter  $H=\dot{a}/a$. For a mode with $\epsilon
=-1,~ \eta_{\epsilon n}>0$ it is proved in \cite{Fos} that there  is
a cosmological initial space-time singularity. One conjectures that
for a mode with $\epsilon =1,~ \eta_{\epsilon n}<0$ there is a
solution of these equations without the cosmological singularity.
The problem of the cosmological singularity is considered in \cite{Ven}.

For the constant Hubble parameter, $H=H_0$, one takes the solution
of the last equation in the form $\psi=\exp (\alpha t)$. Then the
spectrum equation is \be \label{roots-H}
\alpha^2+3H_0\alpha=\epsilon m^2 \,,\ee and we have a deformation of
the mass spectrum.

The Hubble parameter $H_0$ is related with the cosmological constant $\Lambda$ as
$H_0=\sqrt{\Lambda /6}$. If the zeta-function field theory would be a fundamental theory
then we obtain a relation between the Riemann zeros and the cosmological constant. This
gives an additional support to the proposal \cite{Vol2} that the Beilinson conjectures
\cite{Beilinson-conjectures} on the values of $L$-functions of motives can be interpreted
as dealing with the cosmological constant problem.

\section{The Zeta-Function Quantum Field Theory}
To quantize the zeta-function classical field $\phi (x)$ which
satisfies the equation in the Minkowski space \be \label{QFT1}
F(\Box)\phi (x)=0 \,,\ee where $F(\Box)=\xi (\frac{1}{2}+i\Box)$ we can
try to interpret $\phi (x)$ as an operator valued distribution in a
 space $\cal{H}$ which satisfies the  equation (\ref{QFT1}). We
suppose that there is a representation of the Poincare group and a
 vacuum vector $|0\rangle$ in $\cal{H}$ though the space $\cal{H}$
 might be equipped    with indefinite metric and the Lorentz invariance
 can be violated. The Wightman function
$$
W(x-y)=\langle 0|\phi (x)\phi (y)|0\rangle
$$
is a solution of the equation \be \label{QFT2} F(\Box) W (x)=0 \,.\ee
By using Proposition we can write the formal Kallen-Lehmann
\cite{BogSchir} representation
$$
W(x)=\sum_{\epsilon n}\int e^{ixk}f_{\epsilon n} (k)
\delta(k^2+\epsilon m^2_n)dk \,.
$$
A mathematical meaning of this formal expression requires a further investigation.

Quantization of the fields $\psi_{\epsilon n}$ with the Lagrangian (\ref{Z1aa}) can be
performed straightforwardly. We will obtain ordinary scalar fields as well as ghosts and
tachyons. Remind that tachyon presents in the Veneziano amplitude. It was removed by
using supersymmetry. In  Section 6 we shall discuss an approach of how to use a Galois
group and quantum $L$-functions instead of supersymmetry to improve the spectrum.

\section{Quantization of the Riemann-Siegel Function}

One introduces also another  useful function  \be \label{R4}
Z(\tau)=\pi^{-i\tau/2}\frac{\Gamma(\frac{1}{4}+\frac{i\tau}{2})}
{|\Gamma(\frac{1}{4}+\frac{i\tau}{2})|}\zeta(\frac{1}{2}+i\tau)=e^{i\vartheta
(\tau)}\zeta(\frac{1}{2}+i\tau) \,. \ee Here $\Gamma (z)$ is the gamma function.
The function
$Z(\tau)$
 is called the Riemann-Siegel (or Hardy) function \cite{anal-number-theory}.
 It is known that
$Z(\tau)$ is real for real $\tau$ and  there is a bound
 \be \label{R7} Z(\tau)=
 O(|\tau|^{\epsilon}),~~\epsilon>0 \,.\ee
Values $g_n$ such that \be\label{R7b}\vartheta (g_n)=\pi n,~~n=0,1,...\ee
 are known as Gram points. Gram's empirical "law"
 is the tendency for zeros
 of the Riemann-Siegel function  to alternate with Gram points:
\be \label{R7g} (-1)^nZ(g_n)>0\,. \ee An excellent approximation for the Gram point is
given by the formula \be \label{R7gg} g_n\simeq 2\pi\exp[1+W(\frac{8n+1}{8e})]\,,
 \ee where
$W$ is the Lambert function, the inverse function of $f(W)=We^W.$
Note that the Lambert function  appears also in the consideration of the spectrum of
the nonlocal cosmological model in \cite{AJKV}. We have the asymptotic behaviour
$$
W(x)\sim \log x-\log\log x,~~x\to\infty\,,
$$
hence from (\ref{R7gg}) we get
$$
g_n\sim 2\pi n/\log n
$$
(compare with the asymptotic behaviour of the Riemann zeros (\ref{Z2hs})).

One can introduce a natural field theory related with the real valued functions
 $Z(\tau)$ defined by means of the
zeta-function by considering the following Lagrangian
 $$ {\cal L} =   \phi Z({\Box}) \phi \,.
 $$
The integral (\ref{Z2}) converges if $\phi(x)$ is a
 decreasing function
 since  there is the
bound (\ref{R7}).

A generalization of the Riemann zeta-function is studied in \cite{RV}.

It would be interesting also to study the corresponding  "heat equation"
$$
\frac{\partial u}{\partial t}=F(\Delta)u\,,
$$
where $\Delta$ is the Laplace operator. The $p$-adic heat equation
has been considered in \cite{Koz}.

\section{Quantum  $L$-functions}

In this section we briefly discuss $L$-functions and their
quantization. The role of (super) symmetry group here is played by
the Galois group, see \cite{Man,Var,Vol2}.

For any character $\chi$ to modulus $q$ one defines the
corresponding Dirichlet $L$-function \cite{anal-number-theory} by
setting
$$
L(s,\chi)=\sum_{n=1}^{\infty}\frac{\chi (n)}{n^s},~~(\sigma >1) .
$$
If $\chi$ is primitive then $L(s,\chi)$ has an analytic continuation
to the whole complex plane. One introduces the function
$$
\xi (s,\chi)=(\frac{\pi}{k})^{-(s+a)/2}\Gamma
(\frac{s}{2}+\frac{a}{2})L(s,\chi)\,,
$$
where $k$ and $a$ are some parameters. The $\xi$-function is an
entire function and it has a representation similar to (\ref{R6})
\be \label{R6f}
\xi(s,\chi)=Ae^{Bs}s^k\prod_{\rho}(1-\frac{s}{\rho})e^{s/\rho} \,,\ee
where $A,B$ and $k$ are constants. The zeros $\rho$ lie in the
critical strip and symmetrically distributed about the critical line
$\sigma =1/2.$ It is important to notice that unless $\chi$ is real
the zeros will not necessary be symmetric about the real line.
Therefore if we quantize the $L$- function by considering the
pseudodifferential operator
$$
L(\frac{1}{2} +i\Box, \chi)\,,
$$
then we could avoid the appearance of tachyons
and/or ghosts by choosing an appropriate character $\chi$.

The Taniyama-Shimura conjecture relates elliptic curves and modular
forms. It asserts that if $E$ is an elliptic curve over
$\mathbb{Q}$, then there exists a wight-two cusp form $f$  which can
be expressed as the Fourier series
$$
f(z)=\sum a_n e^{2\pi nz}
$$
with the coefficients $a_n$ depending on the curve $E.$  Such a series is a modular form
if and only if  its Mellin transform, i.e. the Dirichlet $L$-series
$$
L(s,f)=\sum a_n n^{-s}
$$
has a holomorphic extension to the full $s$-plane and satisfies a functional equation.
For the elliptic curve $E$ we obtain the $L$-series $L(s,E)$.
The Taniyama-Shimura
conjecture was proved by Wiles and Taylor for semistable elliptic curves and it implies
Fermat`s Last Theorem \cite{Wil}.

Let $E$ be an elliptic curve defined over $\mathbb{Q},$
and $L(s)=L_K(s,E)$
be the $L$-function of $E$ over the field $K.$  The Birch and
Swinnerton-Dyer conjecture asserts that
$$
ord_{s=1}L(s)=r\,,
$$
where $r$ is  the rank of the group $E(K)$ of points of $E$ defined over $K.$
The quantum $L$-function in this case has the form
$$
L(1+i\Box)=A (i\Box)^r+...
$$
where the leading term is expressed in
terms of the Tate-Shafarevich group.

Wiles`s work can be viewed as establishing connections between the automorphic forms and
the representation theory of the adelic groups and the Galois representations. Therefore
it can be viewed as part of the Langlands program in {\it number theory} and the
representation theory \cite{Lan} (for a recent consideration of the  {\it geometrical}
Langlands program see \cite{WGKF}).

Let $G$ be the Galois group of a Galois extension
of  $\mathbb{Q}$ and $\alpha$ a
representation of $G$ in $\mathbb{C}^n.$  There is the Artin
$L$-function $L(s,\alpha)$
associated with $\alpha.$ Artin conjectured that $L(s,\alpha)$
is entire, when $\alpha$
is irreducible, and moreover it is  automorphic: there exists a modular form $f$ such that
$$
L(s,\alpha)=L(s,f)\,.
$$
Langlands formulated the conjecture that $L(s,\alpha)$ is the $L$-function associated to
an automorphic representation of $GL(n,\mathbb{A})$ where
$\mathbb{A}=\mathbb{R}\times\prod_p \mathbb{Q}_p$ is the ring of adeles of $\mathbb{Q}.$
Here $\mathbb{Q}_p$ is the field of $p$-adic numbers. Let $\pi$ be an automorphic
cuspidal representation of $GL(n,\mathbb{A})$ then there is a $L$-function $L(s,\pi)$
associated to $\pi.$ Langlands conjecture (general reciprocity law) states: Let $K$ be a
finite extension of $\mathbb{Q}$ with Galois group $G$ and $\alpha$ be an irreducible
representation of $G$ in $\mathbb{C}^n.$  Then there exists an automorphic cuspidal
representation $\pi_{\alpha}$ of $GL(n,\mathbb{A}_K)$ such that
$$
L(s,\alpha)=L(s,\pi_{\alpha})\,.
$$
Quantization of the $L$-function can be performed similarly to the
quantization of the Riemann zeta-function discussed above by
considering the corresponding pseudodifferential operator $L(\sigma
+i\Box)$ with some $\sigma.$ One speculates that we can not observe
the structure of space-time at the Planck scale but feel only its
motive \cite{Vol2}.

There is a conjecture that the zeros of $L$-functions are distributed like the
eigenvalues of large random matrices from a gaussian ensemble, see \cite{RZ}. The limit
of large matrices in gauge theory (the master field) is derived in \cite{AV3}.

Investigation of the  mass spectrum and field theories  for quantum
$L$-functions we leave for future work.

\section*{Acknowledgements}

 The authors are grateful to Branko Dragovich for useful comments.
 The work of I.A. and I.V. is supported in part
by INTAS grant 03-51-6346. I.A. is also supported  by RFBR grant
05-01-00758  and Russian President's grant NSh-2052.2003.1 and I.V.
is also supported  by RFBR grant 05-01-00884  and Russian
President's grant NSh-1542.2003.1\,.


\begin{thebibliography}{72}

{\small \bibitem{AV} I.Ya.~Aref'eva,  I.V. Volovich, {\it On the Null Energy Condition
and Cosmology}, hep-th/0612098\,.

\bibitem{p-adic} I.V.~Volovich,  {\it p-Adic String},
 Class.Quant.Grav. \textbf{4} (1987) L83\,;\\
 L.~Brekke, P.G.O.~Freund, M.~Olson, E.~Witten,
\textit{Nonarchimedean String Dynamics}, Nucl. Phys. {\bf B302}
(1988) 365\,;\\
P.H. Frampton,  Ya. Okada, {\it Effective Scalar Field Theory of P-Adic String},
Phys. Rev. \textbf{D37} (1988) 3077--3079\,;\\
 V.S.~Vladimirov, I.V.~Volovich, E.I.~Zelenov,
\textit{p-Adic Analysis and Mathe\-matical Physics}, World Sci.,
Singapore, 1994\,;
\\
 A. Khrennikov, {\it P-Adic Valued Distributions in Mathematical Physics},
 Springer, 1994\,.

\bibitem{Man} Yu.I. Manin, {\it Reflections on Arithmetical Physics},
in: "Poiana Brasov 1987, Proceedings, Conformal invariance and
string theory", Boston: Acad.Press, 1989, pp. 293-303\,.

\bibitem{Var}
V.S. Varadarajan, {\it Arithmetic Quantum Physics: Why, What, and
Whither,}  Selected Topics of p-Adic Mathematical Physics and
Analysis, Proc. V.A. Steklov Inst. Math.,  MAIK Nauka/Interperiodica,
2005, Vol. 245, pp. 273-280\,.




\bibitem{IA} I.Ya. Aref'eva, \textit{Nonlocal String Tachyon as
a Model for Cosmological Dark Energy}, AIP Conf. Proc. {\bf 826}
(2006) 301--311; astro-ph/0410443\,.
\bibitem{Calcagni} G. Calcagni,
{\it Cosmological Tachyon from Cubic String Field
Theory}, JHEP \textbf{05} (2006) 012\,; hep-th/0512259\,.
\bibitem{Cline} N. Barnaby, T. Biswas, J.M.  Cline, {\it $p$-adic Inflation},
hep-th/0612230\,.

\bibitem{adelic}
P.G.O. Freund and E. Witten, {\it Adelic String Amplitudes,}
 Phys.Lett.\textbf{B199}(1987) 191\,;\\
 I.V. Volovich, {\it Harmonic analysis and p-adic strings}, Lett.
Math.Phys. \textbf{16} (1988) 61-67\,;\\
I.Ya.~Aref'eva, B. Dragovich, I.V. Volovich, {\it On the adelic string amplitudes},
Phys.Lett. \textbf{B209} (1988) 445\,;\\
B. Grossman, {\it Adelic Conformal Field Theory},
 Phys.Lett.{\bf B215} (1988) 260, Erratum-ibid. {\bf B219} (1989) 531\,;\\
V.S. Vladimirov, {\it Adelic Formulas for Gamma and Beta Functions of One-Class Quadratic
Fields: Applications to 4-Particle Scattering String Amplitudes},
Proc. Steklov Math. Inst., {\bf 248} (2000), 76-89; math-ph/0004017\,;\\
B. Dragovich, {\it Adelic harmonic oscillator},
Int.J.Mod.Phys. {\bf A10} (1995) 2349-2365;\\ hep-th/0404160\,.



\bibitem{anal-number-theory} E. C. Titchmarsh, {\it The theory of the Riemann
zeta-function}, Oxford, Clarendon Press, 1986\,.\\
 A. A. Karatsuba, S. M. Voronin, {\it The Riemann zeta-function},
  Walter de Gruyter Publishers, Berlin-New York 1992\,.\\
  K. Chandrasekharan, {\it Arithmetic Functions,} Springer, 1972\,.




\bibitem{zeta-mp} L.D. Faddeev and B.S. Pavlov,
{\it Scattering theory and automorphic functions},
Semin. of Steklov Math. Inst. of Leningrad, {\bf 27}
(1972), 161-193\,;\\
B. Julia, {\it Statistical theory of numbers},
in Number Theory and Physics, Springer Proceedings in Physics,
{\bf 47} (1990) p. 276\,;\\
A. Connes. {\it Trace formula in Noncommutative geometry and
the zeroes of the Riemann zeta-function}. Selecta. Math. (NS) {\bf 5}
(1999) 29-106\,;\\
 M.V. Berry and J.P.Keating,
{\it The Riemann zeros and eigenvalue asymptotics},
SIAM Review {\bf 41} (2) 236, 1999\,;\\
C. Castro,
{\it On p-adic stochastic dynamics, supersymmetry and the Riemann conjecture},
Chaos Solitons \& Fractals {\bf 15},  15 (2003); physics/0101104\,;\\
G. Cognola, E. Elizalde and S. Zerbini, {\it Heat-kernel expansion on noncompact domains
and a generalized zeta-function regularization procedure}. J.Math.Phys. {\bf 47},
(2006) 083516\,.

\bibitem{KozVol} V.V. Kozlov, I.V. Volovich, {\it Finite Action Klein-Gordon
    Solutions on Lorentzian Manifolds},
    Int. J. Geom. Meth. Mod. Phys. \textbf{3} (2006) 1349--1358;
    gr-qc/0603111\,;\\
V.V. Kozlov, I.V. Volovich, {\it Mass Spectrum, Actons and
Cosmological Landscape},\\
 hep-th/0612135\,.

\bibitem{Wil} A. Wiles, {\it Modular Elliptic Curves and Fermat's Last Theorem }
Annals of Mathematics, {\bf 141} (1995), 443-551\,;\\
R. Taylor, A. Wiles,  {\it Ring-Theoretic Properties of Certain
Hecke
Algebras}, Annals of Mathematics, {\bf 141} (1995), 553-572\,;\\
Yves Hellegouarch, {\it Invitation to the Mathematics of
 Fermat-Wiles}, Academic Press, 2001\,.

\bibitem{Lan} R.P. Langlands, {\it Problems in the theory of automorphic forms},
in Lect. Notes in Math. 170, pp. 18–61, Springer Verlag, 1970\,.


\bibitem{Hor}
V.P. Maslov, {\it Operator Methods,} Nauka, 1973\,.\\
 L. Hormander, {\it The Analysis of Linear Partial Differential
Operators,  III Pseudo-Differential Operators, IV Fourier Integral
Operators,} Springer-Verlag, 1985\,.\\
M.A. Shubin, {\it Pseudodifferential Operators and Spectral Theory,}
Nauka, 1978.

\bibitem{VlaVolY} V.S. Vladimirov, Ya.I. Volovich,
\textit{Nonlinear Dynamics Equation in p-Adic String Theory}, Theor.
Math. Phys., {\bf 138} (2004) 297; math-ph/0306018\,.


\bibitem{PaisU} A. Pais and G.E. Uhlenbeck,
{\it On Field Theories with Nonlocalized Action},
 Phys.Rev. {\bf 79}: 145-165, 1950\,.

\bibitem{BogSchir} N.N. Bogoliubov and D.V. Schirkov, {\it Introduction to the
Theory Of Quantized Fields}, Springer, 1984\,.


 \bibitem{AJKV}
 I.Ya. Aref'eva, A.S. Koshelev, S.Yu. Vernov, {\it
 Crossing of the $w=-1$ Barrier by $D3$-brane Dark Energy Model},
 Phys.Rev. D72 (2005) 064017\,; astro-ph/0507067 \,;\\
I.Ya.~Aref'eva, A.S.~Koshelev,
 {\it Cosmic Acceleration and Crossing of $w=-1$ barrier
 from Cubic Superstring Field Theory}, hep-th/0605085\,;\\
 A.S.~Koshelev, {\it Non-local SFT Tachyon and Cosmology},
 hep-th/0701103\,;\\
I.Ya.~Aref'eva, L.V. Joukovskaya and S.Yu.Vernov,
{\it Bouncing and Accelerating Solutions  in Nonlocal Stringy Models},
hep-th/0701189\,.

\bibitem{Fos} S. Foster, {\it Scalar Field
Cosmologies and the Initial Space-Time Singularity,}  gr-qc/9806098\,.

\bibitem{Ven}  M. Gasperini, G. Veneziano, {\it The Pre-big bang
scenario in string cosmology},
Phys.Rept. \textbf{373} (2003) 1-212\,; hep-th/0207130\,.

\bibitem{Vol2}
I.V. Volovich, {\it D-branes, Black Holes and $SU(\infty)$ Gauge Theory,}
hep-th/9608137\,;\\
I.V. Volovich, {\it From p-adic strings to etale strings}, Proc.
Steklov Math. Inst, {\bf 203} (1994)41-48\,.

\bibitem{Beilinson-conjectures} M Rapoport, N Schappacher, P Schneider,
 {\it Beilinson's conjectures on special values of L-functions},
 Academic Press, 1988\,.


\bibitem{RV}
K. Ramachandra and I. V. Volovich, {\it A generalization of the
Riemann zeta-function}, Proc. Indian Acad. Sci. (Math. Sci.) {\bf
99} (1989) 155-162\,.

\bibitem{Koz} V.A. Avetisov, A.H. Bikulov, S.V. Kozyrev, V.A. Osipov,
{\it p-Adic Models of Ultrametric Diffusion Constrained by
Hierarchical Energy Landscapes,} J.Phys. A: Math. Gen., {\bf 35}
(2002) 177-189;
cond-mat/0106506;\\
S.V. Kozyrev, {\it p-Adic pseudodifferential operators and p-adic wavelets,}
math-ph/0303045\,.



\bibitem{WGKF}
 A. Kapustin and E. Witten, {\it  Electric-Magnetic Duality And The Geometric
Langlands \\ Program}, hep-th/0604151;\\
S. Gukov and E. Witten, {\it Gauge Theory, Ramification, And The
Geometric Langlands \\ Program}, hep-th/0612073;\\
 E. Frenkel, {\it Lectures on the Langlands Program and Conformal
Field Theory}, hep-th/0512172\,.

\bibitem{RZ} N.M. Katz and P. Sarnak, {\it Zeroes of zeta-functions and symmetry},
Bull. Amer. Math. Soc. (N.Y.) {\bf 36} (1999),1-26;\\
J.P. Keating and N.C. Snaith, {\it Random matrix theory and
$L$-functions at $s=1/2$}, Comm.Math. Phys. {\bf 214} (2000),
91-110\,.
\bibitem{AV3}
I.Ya. Aref'eva and I.V. Volovich, {\it The Master Field for QCD and
$q$-Deformed Quantum Field Theory}, Nucl.Phys. {\bf B 462} (1996)
600-612; hep-th/9510210\,.

}
\end{thebibliography}
 \end{document}